\documentclass[apj]{emulateapj}
\usepackage{mathptmx}
\newcommand  \kms      {\ifmmode {\rm km\,s}^{-1} \else km\,s$^{-1}$\fi}

\newcommand  \ergs     {\ifmmode {\rm ergs\,s}^{-1} \else ergs s$^{-1}$\fi}
\newcommand  \ergcms   {\ifmmode {\rm ergs\,cm}^{-2}\,{\rm s}^{-1}
                        \else ergs\,cm$^{-2}$\,s$^{-1}$\fi}
\newcommand  \ergcmsA {\ifmmode{\rm ergs\,cm}^{-2}\,{\rm s}^{-1}\,{\rm\AA}^{-1}
                        \else ergs\,cm$^{-2}$\,s$^{-1}$\,\AA$^{-1}$\fi}
\newcommand \ergcmsHz {\ifmmode{\rm ergs\,cm}^{-2}\,{\rm s}^{-1}\,{\rm Hz}^{-1}
                        \else ergs\,cm$^{-2}$\,s$^{-1}$\,Hz$^{-1}$\fi}
\newcommand  \phcms    {\ifmmode {\rm ph\,cm}^{-2}\,{\rm s}^{-1}
                        \else ,ph\,cm$^{-2}$\,s$^{-1}$\fi}
\newcommand  \phcmsA   {\ifmmode {\rm ph\,cm}^{-2}\,{\rm s}^{-1}\,{\rm\AA}^{-1}
                        \else ph\,cm$^{-2}$\,s$^{-1}$\,\AA$^{-1}$\fi}
\newcommand{\mbh}{$M_{\rm BH}$}
\def\micron{\ifmmode \mu{\rm m} \else $\mu$m\fi}
\def\kms{\ifmmode {\rm km\,s}^{-1} \else km\,s$^{-1}$\fi}
\def\Hubble{\ifmmode {\rm km\,s}^{-1}\,{\rm Mpc}^{-1}
        \else km\,s$^{-1}$\,Mpc$^{-1}$\fi}
\def\ergsec{\ifmmode {\rm ergs\;s}^{-1} \else ergs s$^{-1}$\fi}
\def\ergscm{\ifmmode {\rm ergs\,s}^{-1}\,{\rm cm}^{-2}
          \else ergs\,s$^{-1}$\,cm$^{-2}$\fi}
\def\ergscmA{\ifmmode {\rm ergs\,s}^{-1}\,{\rm cm}^{-2}\,{\rm \AA}^{-1}
          \else ergs\,s$^{-1}$\,cm$^{-2}$\,\AA$^{-1}$\fi}
\def\ergscmHz{\ifmmode {\rm ergs\,s}^{-1}\,{\rm cm}^{-2}\,{\rm Hz}^{-1}
          \else ergs\,s$^{-1}$\,cm$^{-2}$\,Hz$^{-1}$\fi}
\def\Msun{\ifmmode M_{\odot} \else $M_{\odot}$\fi}
\def\Lsun{\ifmmode L_{\odot} \else $L_{\odot}$\fi}
\def\qo{\ifmmode q_{0} \else $q_{0}$\fi}
\def\Ho{\ifmmode H_{0} \else $H_{0}$\fi}
\def\ho{\ifmmode h_{0} \else $h_{0}$\fi}
\def\qo{\ifmmode q_{0} \else $q_{0}$\fi}
\def\ao{\ifmmode a_{0} \else $a_{0}$\fi}
\def\to{\ifmmode t_{0} \else $t_{0}$\fi}
\def\ltsim{\raisebox{-.5ex}{$\;\stackrel{<}{\sim}\;$}}
\def\gtsim{\raisebox{-.5ex}{$\;\stackrel{>}{\sim}\;$}}
\def\Halpha{\ifmmode {\rm H}\alpha \else H$\alpha$\fi}
\def\Hbeta{\ifmmode {\rm H}\beta \else H$\beta$\fi}
\def\hb{\ifmmode {\rm H}\beta \else H$\beta$\fi}
\def\Hgamma{\ifmmode {\rm H}\gamma \else H$\gamma$\fi}
\def\Hdelta{\ifmmode {\rm H}\delta \else H$\delta$\fi}
\def\Lya{\ifmmode {\rm Ly}\alpha \else Ly$\alpha$\fi}
\def\Lyb{\ifmmode {\rm Ly}\beta \else Ly$\beta$\fi}
\def\hi{\ifmmode \mbox{{\rm H}\,{\sc i}} \else H\,{\sc i}\fi}

\def\ciii{\ifmmode {\rm C}\,{\sc iii} \else C\,{\sc iii}\fi}
\def\civ{C\,{\sc iv}\,$\lambda1549$}

\def\oiii{[O\,{\sc iii}]\,$\lambda5007$}

\def\o5007{[O\,{\sc iii}]\,$\lambda5007$}

\def\ne212m {[Ne\,{\sc ii}]\,$12.8 \mu m$}

\def \Lop{$L_{5100}$}
\def \Ledd{$L/L_{\rm Edd}$}

\def  \kms         {\hbox{km s$^{-1}$}}          % kilometers per sec
\def  \ergs        {\hbox{ergs s$^{-1}$}}              % erg/sec
\def  \La          {\ifmmode {\rm Ly}\alpha \else Ly$\alpha$\fi}
\def  \Ka          {\ifmmode {\rm K}\alpha \else K$\alpha$\fi}
\def  \Lb          {\ifmmode {\rm L}\beta \else L$\beta$\fi}
\def  \Ha          {\ifmmode {\rm H}\alpha \else H$\alpha$\fi}
\def  \Hb          {\ifmmode {\rm H}\beta \else H$\beta$\fi}
\def  \Pa          {\ifmmode {\rm P}\alpha \else P$\alpha$\fi}
\def  \CIIIb       {\ifmmode {\rm C}\,{\sc iii]}\,\lambda1909
                     \else C\,{\sc iii]}\,$\lambda1909$\fi}
\def  \CIV         {\ifmmode {\rm C}\,{\sc iv}\,\lambda1549
                     \else C\,{\sc iv}\,$\lambda1549$\fi}
\def  \MgII         {\ifmmode {\rm Mg}\,{\sc ii}\,\lambda2798
                     \else Mg\,{\sc ii}\,$\lambda2798$\fi}
\def  \OVI         {\ifmmode {\rm O}\,{\sc vi}\,\lambda1035
x
                     \else O\,{\sc vi}\,$\lambda1035$\fi}

\journalinfo{The Astrophysical Journal, ???:???--???, 200? ?????? ??}
\slugcomment{Received 2007 May 17; accepted 2007 August 27}

\shorttitle{BLACK-HOLE MASS AND GROWTH RATE}
\shortauthors{NETZER ET AL.}

\begin{document}
%%%%%%%%%%%%%%%%%%%%%%%%%%

\title{Black-Hole Mass and Growth Rate at High Redshift}

\author{
Hagai Netzer,\altaffilmark{1}
Paulina Lira,\altaffilmark{2}
Benny Trakhtenbrot,\altaffilmark{1}
Ohad Shemmer,\altaffilmark{3}
and Iara Cury\altaffilmark{4}
}
\altaffiltext{1} {School of Physics and Astronomy and the Wise
  Observatory, The Raymond and Beverly Sackler Faculty of Exact
  Sciences, Tel-Aviv University, Tel-Aviv 69978, Israel}
\altaffiltext{2} {Departamento de Astronom\'ia, Universidad de Chile,
  Camino del Observatorio 1515, Santiago, Chile}
\altaffiltext{3} {Department of Astronomy and Astrophysics, 525 Davey
  Laboratory, Pennsylvania State University, University Park, PA
  16802, USA}
\altaffiltext{4} {Astronomy Department Yale University, P.O. Box
  208101, New Haven, CT 06520-8101, USA}

\begin{abstract}
  We present new $H$ and $K$ bands spectroscopy of 15 high luminosity
  active galactic nuclei (AGNs) at redshifts 2.3--3.4 obtained on
  Gemini South.  We combined the data with spectra of additional 29
  high-luminosity sources to obtain a sample with $10^{45.2}<\lambda
  L_{\lambda}(5100\mbox{\AA})<10^{47.3}$ \ergs\ and black hole (BH)
  mass range, using reverberation mapping relationships based on the
  \hb\ method, of $10^{8.8} - 10^{10.7}$ \Msun. We do not find a
  correlation of \Ledd\ with \mbh\ but find a correlation with
  $\lambda L_{\lambda}(5100\mbox{\AA})$ which might be due to
  selection effects. The \Ledd\ distribution is broad and covers the
  range $\sim$0.07--1.6, similar to what is observed in lower
  redshift, lower luminosity AGNs. We suggest that this consistently
  measured and calibrated sample gives the best representation of
  \Ledd\ at those redshifts and note potential discrepancies with
  recent theoretical and observational studies.  The lower accretion
  rates are not in accord with growth scenarios for BHs at such
  redshifts and the growth times of many of the sources are longer
  than the age of the universe at the corresponding epochs. This
  suggests earlier episodes of faster growth at $z\gtsim3$ for those
  sources. The use of the \civ\ method gives considerably different
  results and a larger scatter; this method seems to be a poor \mbh\
  and \Ledd\ estimator at very high luminosity.
\end{abstract}

\section{Introduction}
\label{introduction}

Studies of black-hole (BH) growth at various redshifts, and the
comparison with galaxy evolution and star formation, has been a very
active area of research for several years. In particular, there are
several suggestions that very massive BHs grew faster at early epochs
while the growth of less massive BHs extends over longer periods and
is significant even at $z=0$. For example, Marconi et al. (2004) used
the \hbox{X-ray} luminosity function of active galactic nuclei (AGNs)
to suggest a specific growth pattern as a function of cosmic time.
According to these authors, BHs with \mbh$\gtsim10^{8}$ \Msun\
attained 50\% of their mass by $z=2$ and 90\% of their mass by $z=1$
(Marconi et al. 2004, Fig. 2). Smaller BHs grew slower at earlier
times and many active BHs with \mbh\,$\sim 10^7$ \Msun\ are still
growing today. Similar scenarios, under the general terminology of
``anti-hierarchical growth of supermassive BHs'', have been presented
by Merloni (2004) and others. Those studies assume that the growth
rate of very massive BHs at high redshifts approached the Eddington
limit. More recent studies, e.g., by Volonteri et al. (2006) and
Hopkins et al. (2006), focus on the importance of the Eddington ratio,
\Ledd, in determining BH evolution at all redshifts. This includes
also hierarchical models for the evolution of the most massive
BHs. All these models can be tested observationally by direct
measurements of BH mass and accretion rate at high redshifts, provided
high quality observations and reliable methods for determining \mbh\
are available.

Current BH mass estimates are based on reverberation mapping that
provides a way to measure the emissivity-weighted size of the
broad-line region (BLR) in type-I AGNs as a function of the optical
continuum luminosity ($\lambda L_{\lambda}$ at 5100\AA, hereafter
\Lop; see Kaspi et al. 2000; Kaspi et al. 2005, hereafter K05 and
references therein). This scaling has been used to obtain a
``single-epoch'' estimate of the BH mass by combining the BLR size
derived from \Lop\ with a measure of the gas velocity obtained from
FWHM(\hb) (hereafter ``the \hb\ method'').  There are obvious
limitations to this method due to the somewhat vague definition of the
BLR size, variable source luminosity, BLR geometry, and somewhat
uncertain line widths. There are also questions regarding the exact
slope of the BLR-size vs. the source luminosity (e.g. Bentz et
al. 2006).  These translate to a factor of $\sim 2$ uncertainty on the
derived masses. An additional uncertainty is associated with the
limited luminosity range of the K05 sample and hence the need to
extrapolate the relationship beyond its highest luminosity end, at
\Lop$\simeq 2 \times 10^{46}$ \ergs\ (see, however, the new result of
Kaspi et al. 2007).

Other combinations of continuum luminosity and line widths have also
been used, especially in the study of high-redshift sources (e.g.,
McLure \& Dunlop 2004; Vestergaard 2004; Vestergaard \& Peterson
2006). These seem to be associated with a larger uncertainty on the
derived masses especially when the \civ\ line width, in combination
with $\lambda L_{\lambda}$ at $\sim 1400$\,\AA\ (hereafter ``the \civ\
method'') are used (e.g., Baskin and Laor 2005). A way to extend the
\hb\ method to high redshifts is by high quality near-infrared
spectroscopy. Measurements in a moderate size sample (29 sources) of
this type are presented in Shemmer et al. (2004, hereafter S04) who
used $H$- and $K$-band spectroscopy for obtaining single-epoch BH
masses and accretion rates in very high luminosity AGNs. More recent
work of this type (a sample of 9 sources with $1.08 < z < 2.32$) is
reported in Sulentic et al. (2006).

\begin{deluxetable*}{lccclc}
\tablecolumns{6}
\tablewidth{0pt}
\tablecaption{GNIRS Observation Log \label{table_log}}
\tablehead{
\colhead{Object ID (SDSS~J)} &
\colhead{$z_{\rm sys}$\tablenotemark{a}} &
\colhead{$z_{\rm SDSS}$\tablenotemark{b}} &
\colhead{Band} &
\colhead{Obs. Date} &
\colhead{Exp. Time (hr)}
}
\startdata
025438.37$+$002132.8 & 2.456 & 2.463 & $H$ & 2005 Dec 7 & 1.0 \\ % =600x6  6
083630.55$+$062044.8 & 3.397 & 3.397 & $K$ & 2005 Nov 27 & 1.0 \\ % =450x8  9
095141.33$+$013259.5 & 2.411 & 2.429 & $H$ & 2006 Jan 24 & 2.5 \\ % =600x15 15
100710.70$+$042119.1 & 2.363 & 2.363 & $H$ & 2006 Mar 26 & 2.0 \\ % =600x12 12
101257.52$+$025933.2 & 2.434 & 2.434 & $H$ & 2006 Feb 08 & 0.5 \\ % =600x3  3
105511.99$+$020751.9 & 3.391 & 3.384 & $K$ & 2005 Apr 21 & 1.0 \\ % =400x9  9
113838.26$-$020607.2 & 3.352 & 3.343 & $K$ & 2005 Apr 21 & 1.0 \\ % =400x9  9
115111.20$+$034048.3 & 2.337 & 2.337 & $H$ & 2006 Apr 16,25 & 2.5 \\ %=600x15 15
115304.62$+$035951.5 & 3.426 & 3.432 & $K$ & 2005 Apr 23  & 0.7 \\ % =400x6  6
115935.64$+$042420.0 & 3.451 & 3.448 & $K$ & 2005 Apr 23  & 0.7 \\ % =400x6  6
125034.41$-$010510.5 & 2.397 & 2.397 & $H$ & 2006 Apr 26,27 & 2.7 \\ %=600x16 15
144245.66$-$024250.1 & 2.356 & 2.343 & $H$ & 2006 Jul 19  & 1.0 \\ % =600x6  6
153725.36$-$014650.3 & 3.452 & 3.452 & $K$ & 2005 Apr 21  & 0.7 \\ % =400x6  6
210258.21$+$002023.4 & 3.328 & 3.343 & $K$ & 2006 Jul 19  & 1.3 \\ % =400x12 12
210311.69$-$060059.4 & 3.336 & 3.336 & $K$ & 2005 May 31 &
\phm{1}0.7 % =400x6 6
\enddata
\tablenotetext{a}{Systemic redshift measured from the [\ion{O}{3}]
  lines.}
\tablenotetext{b}{Redshift obtained from the SDSS archive, based on
  rest-frame UV emission lines.}
\end{deluxetable*}

This paper presents BH masses and accretion rates obtained with the
\hb\ method for a new sample of 15 sources at two redshift bands
around $z\simeq2.3$ and $z\simeq3.4$. Together with our earlier work
(S04) we can now use the \hb\ method to investigate the mass and
accretion rate of high redshift sources over a range of 2.5 dex in
luminosity. In \S~\ref{observations} we describe the observations and
their analysis, and in \S~\ref{discussion} we present the main results
and a discussion of our findings including an assessment of the growth
rate of high-redshift AGNs, and the evaluation of BH mass measurements
obtained with the \civ\ method as a replacement for the \hb\ method at
high redshifts.

\section{Observations, Data Reduction, and Mass Determination}
\label{observations}

The sample described in this paper contains 15 high redshift,
high-luminosity AGNs that were selected by their luminosity and
redshift. The redshift range stems from the requirement to directly
measure the \hb\ line and the 5100\,\AA\ continuum from the ground in
order to use the \hb\ method, which we consider to provide the most
reliable BH mass estimates. The redshift ranges are dictated by the
wavelengths of the $H$ and $K$ bands and are $\sim$2.1--2.5 and
$\sim$3.2--3.5, respectively. The luminosity is dictated by the goal
of going down the AGN luminosity function, starting from its top, and
measuring masses and accretion rates of fainter and fainter
sources. Given this, we chose sources that are 5--10 times less
luminous than the mean luminosity in the S04 sample. For the basic
sample we chose the Sloan Digital Sky Survey (SDSS; e.g., York et
al. 2000) which contains flux calibrated data and hence luminosity
estimates for all sources.

Spectroscopic observations were obtained with the Gemini Near Infrared
Spectrograph (GNIRS) on Gemini South under programs GS-2005B-Q-28,
GS-2006A-Q-58 and GS-2005A-Q-51. The long slit observations were
acquired in the $H$ or $K$ bands depending on the redshift of the
source. The slit width was 1\arcsec\ and the targets were nodded along
the slit to obtain a good background subtraction. The 32\,l/mm grating
was used in all observations resulting in $R\sim 640$ and 850, $\Delta
\lambda \sim 1.46-1.89 \, \mu$m and $\Delta \lambda \sim 1.83-2.49 \,
\mu$m for the $H$ and $K$ bands, respectively. Typical exposure times
of the sub-integrations were 400 to 600 seconds. More details are
given in Table~\ref{table_log} where we list all 15 sources and assign
to each the systemic redshift measured from their \oiii\ lines. These
can be somewhat different from the SDSS redshifts that were obtained
by measuring rest-frame ultraviolet (UV) broad-emission lines and are
also listed in the table.

The reduction of the raw spectroscopic data was done using the {\sc
  gemini} package in IRAF. The pipeline combines observations from
different nodded positions to obtain background-subtracted images,
determines the wavelength calibration, registers the frames, and
produces a final averaged image. Extraction of the spectra and flux
calibration were performed using standard IRAF tasks. Special care was
taken to correct for telluric absorption. This was done by observing
early-type stars right before or after the science targets and at
similar~air~masses.

To obtain a more accurate flux calibration of our spectra, $H$- and
$K$-band photometry was obtained for 13 of our sources using the ISPI
detector on the Cerro Tololo Inter-American Observatory 4\,m telescope
on 2007 February 12. The data were reduced in the standard way using
the {\sc xdimsun} package in IRAF\footnote{IRAF (Image Reduction and
  Analysis Facility) is distributed by the National Optical Astronomy
  Observatories, which are operated by the Association of Universities
  for Research in Astronomy, Inc., under cooperative agreement with
  the National Science Foundation.}. The calibration was achieved
using Two Micron All Sky Survey
(2MASS\footnote{http://www.ipac.caltech.edu/2mass}) stars in the
$10\arcmin \times 10\arcmin$ field of view of the instrument. Only
stars with good quality flags in the 2MASS All-Sky Catalog of Point
Sources (Cutri et al. 2003) were used, resulting in a $\sim 10\%$
accuracy. All luminosities listed below are based on these values
except for the two sources that were not observed where we use the
flux from the spectroscopy. The typical RMS difference between the two
methods is about 15\%.

\begin{figure*}
\vspace{-0.9in}
\epsscale{1.3}
\plotone{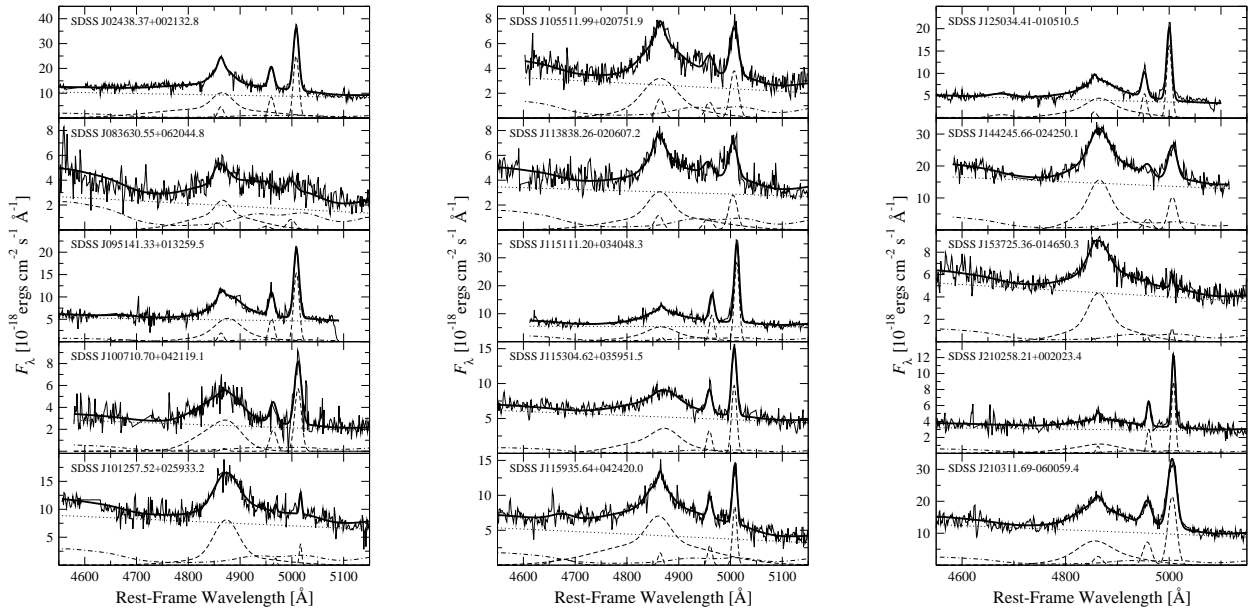}
\caption{GNIRS spectra of our sample of 15 high-redshift AGNs. The
  best-fit model (thick solid curve) in each panel is composed of a
  continuum component (dotted curve), an Fe~{\sc ii}-emission complex
  (thin dot-dashed curve), and H$\beta$ and [O~{\sc iii}] emission
  lines (thick dashed curve).}
\label{fig:spectra_1}
\end{figure*}

We applied a fitting procedure similar to the one described in Netzer
\& Trakhtenbrot (2007; hereafter NT07) to fit all spectra with the
various components expected in this range. In short, we fit a linear
continuum between rest-frame $\sim 4700$\AA\ and $\sim 5100$\AA\ and
then a five-component Gaussian emission line model to the
continuum-subtracted spectrum: two components for the broad \hb\ line,
one for the narrow \hb\ line, and two for the
[\ion{O}{3}]\,$\lambda\lambda 4959,5007$ \AA\ lines. This fit serves
to obtain a first estimate of FWHM(\hb). We then use the Boroson \&
Green (1992) \ion{Fe}{2} template, convolved with a single Gaussian
with the above FWHM (i.e. the one obtained from the combination of the
two broad components), to fit the \ion{Fe}{2} emission complex over
the range 4400--4650\AA. This fit, extended to the entire wavelength
range by using the template, is used to estimate the \ion{Fe}{2} line
contributions to the continuum bands and to improve the continuum
definition. Using the initial \ion{Fe}{2} template, we obtain a
modified continuum-subtracted spectrum and repeat the line fitting
process. A second iteration \ion{Fe}{2} model is obtained and then
subtracted from the spectrum, and a final, five-component Gaussian fit
of the \hb\ and [\ion{O}{3}] lines is performed. In the final stage,
the two BLR Gaussians are limited to 1,500$<$FWHM$<$20,000 \kms\ and
the NLR components are forced to have the same FWHM for all three
lines. This component is not allowed to exceed 1,200 \kms, which is
smaller than the FWHM([\ion{O}{3}]) values of some of the S04 sources
(see Table~1 of Netzer et al. 2004) but appears to be adequate for the
present sample. The monochromatic luminosity at 5100\AA, \Lop, is
measured directly from the fitted continuum.  The results of our
line-fitting procedure are given in Fig.~\ref{fig:spectra_1} where we
show the observed spectrum, the model components, and the final fit
for all 15 sources.

The measurement of \mbh\ is a crucial point and requires some
discussion. The K05 study suggests that the correlation of BLR size as
deduced from the \hb\ lag with respect to the optical continuum is
given by $R_{\rm BLR} \propto L^{\alpha}$ where
\hbox{$\alpha=0.65-0.7$}.  This slope is obtained from fitting 34
sources covering about four orders of magnitude in \Lop. More recent
work by Bentz et al. (2006) used an improved stellar subtraction
procedure for several of the lower luminosity sources in the K05
sample which reduced \Lop\ for those cases. Using 26 sources they
obtained $\alpha=0.52$. A more recent study by the same group (Bentz
et al. 2007) that includes four additional sources (but again not the
entire K05 sample) with improved stellar subtraction gives
$\alpha=0.54 \pm 0.04$.

Given the very large luminosity range (about a factor \hbox{$2 \times
  10^4$} in \Lop) of the K05 sample, it is not at all clear that the
slopes at low and high luminosities are the same. It is thus justified
to use the higher luminosity sources to obtain the most appropriate
slope for extrapolating to luminosities larger than those of the K05
sample. Similarly, we suggest to use only the lower luminosity sources
in K05 (after including the Bentz et al. 2007 corrections) when
looking for the best extrapolation to very low luminosities. This
approach was used by NT07 in their study of the SDSS sample. Since
most of the source analyzed here have extremely large \Lop\
($10^{45.2-47.3}$ \ergs), we chose to adopt the same approach and
obtain a best slope by fitting only those sources in K05 with
\Lop$>10^{43.5}$~\ergs.  This gives the same expression as used by
NT07,
\begin{equation}
  M_{\rm BH}=1.05\times 10^8 \left [\frac{L_{5100}}{10^{46}\,
      {\rm ergs\,s^{-1}}}\right]^{0.65} \left[\frac
    {{\rm FWHM}({\rm H}\beta)}{10^3 \,{\rm km\,s^{-1}}}\right]^2 \,\, \Msun \, .
\label{eq:M_L}
\end{equation}
For comparison, we also calculated \mbh\ and accretion rate using the
expression given in Bentz et al. (2007 Eq. 2). This gives masses that
are smaller by a factor of $\sim 1.2$ for our lowest \Lop\ sources,
and by a factor of $\sim 1.9$ for our highest luminosity AGNs. These
alternative estimates are compared below with the results obtained by
using Eq.~\ref{eq:M_L}.

The calculation of $L_{\rm bol}/L_{\rm Edd}$ (hereafter \Ledd) is
based on a bolometric correction, $f_L$, which is somewhat luminosity
dependent, and for the sample in hand is of order 5--7 (NT07 and
references therein). For most applications described here we assumed
$f_L=7$ [i.e., \hbox{\Ledd$=7\times$\Lop$/(1.5 \times
  10^{38}$\,\mbh/\Msun)}]. Marconi et al. (2004) give a specific
expression for $f_L$ that translates to bolometric correction factors
of 5.4--6.4 for our sample (see also Richards et al. 2006 on this
issue but note the different method of counting the IR luminosity in
that paper). We have carried the analysis described below using those
values but given the small range in $f_L$, the conclusions hardly
change. Table~\ref{table_properties} shows the results obtained from
the above measurements and fits by assuming a standard cosmology with
$\Omega_{\Lambda}=0.7$, $\Omega_{m}=0.3$, and $H_{0}=70$ \kms\
Mpc$^{-1}$.

\begin{deluxetable*}{lcccccccc}
\tablecolumns{9}
\tablewidth{0pt}
\tablecaption{Observed and Derived Properties
  \label{table_properties} }
\tablehead{
  \colhead{Object ID (SDSS~J)}  &
  \colhead{$z_{\rm sys}$}  &
  \colhead{$\log$ \Lop} &
  \colhead{$\log$ $\lambda L_{\lambda}$(1450\AA)}  &
  \colhead{FWHM(\hb)} &
  \colhead{FWHM(C~{\sc iv})}   &
  \colhead{$\log$ \mbh} &
  \colhead{$\log$ \Ledd} &
  \colhead{$t_{\rm grow}$\tablenotemark{a} / } \\
   & & (\ergs) & (\ergs) & (\kms) & (\kms) & (\Msun) & & $t_{\rm
    universe}$ }
\startdata
025438.37$+$002132.8 & 2.456& 45.85 & 45.93 &4164 & 4753 & 9.162 &
-0.64$\pm$0.08 & 0.8\\
083630.55$+$062044.8 & 3.397& 45.53 & 46.12 &3950 & 5878 & 8.909 &
-0.71$\pm$0.15 & 1.3 \\
095141.33$+$013259.5 & 2.411& 45.55 & 45.90 &4297 & 5289 & 8.997 &
-0.78$\pm$0.15 & 1.1 \\
100710.70$+$042119.1 & 2.363& 45.17& 45.67 & 5516 & 5495 & 8.96 &
-1.13$\pm$0.15  &2.4 \\
101257.52$+$025933.2 & 2.434& 45.73& 45.95 & 3892 & 5862 & 9.029 &
-0.62$\pm$0.08 & 0.8 \\
105511.99$+$020751.9 & 3.391 & 45.70 & 46.24 & 5424 & 5476 & 9.294 &
-0.93$\pm$0.08 & 2.4 \\
113838.26$-$020607.2 &3.352& 45.79 &46.18 &4562 &6098 &9.271 
& -0.74$\pm$0.15 &1.5\\
115111.20$+$034048.3 & 2.337& 45.58 & 45.58 & 5146 & 2860 & 9.171 &
-0.92$\pm$0.08 & 1.5 \\
115304.62$+$035951.5 &3.426& 46.04 &46.37 &5521 &1773 &9.529 & 
-0.82$\pm$0.08 &2.0\\
115935.64$+$042420.0 &3.451& 45.92 &46.43 &5557 & 4160 & 9.460 &
-0.89$\pm$0.15 & 2.2\\
125034.41$-$010510.5 & 2.397& 45.41 &45.71 &5149 & 4234 & 9.061 &
-0.98$\pm$0.08 & 1.8\\
144245.66$-$024250.1 & 2.356& 46.03 & 45.90 & 3661 & 3277 & 9.166 &
-0.47$\pm$0.08 & 0.6\\
153725.36$-$014650.3 & 3.452& 45.98 & 46.44 &3656 & 5650 & 9.133 &
-0.49$\pm$0.08 & 0.9\\
210258.21$+$002023.4 & 3.328& 45.79 & 45.86 & 7198& 2355 & 9.599 &
-1.14$\pm$0.08 & 4.0\\
210311.69$-$060059.4 & 3.336& 46.30 & 46.24 &6075 & 4951 & 9.785 &
-0.81$\pm$0.08 & \phm{}2.0
\enddata
\tablenotetext{a}{Assuming $\eta=0.1$, $f_L=7$, $M_{\rm
    seed}=10^4$\,\Msun, and $f_{\rm active}=1$.}
\end{deluxetable*}

All the 29 spectra of the S04 sample were refitted using a similar
procedure and BH masses were recalculated using Eq.~\ref{eq:M_L}
(which is slightly different from the one used in S04). Some of the
S04 data are of poorer quality compared to the new GNIRS observations
which resulted in larger uncertainties. In addition, slight
mis-centering of the \hb\ line in the $H$ or $K$ bands forced us, in
several cases, to perform the \ion{Fe}{2} fit on the \ion{Fe}{2}
complex longword of the \hb\ line. Another difference from the S04
procedure is the inclusion of the narrow \hb\ component in the
fits. As a result, several of the newly measured FWHM(\hb) values are
somewhat larger than those found by S04. Given all this, we consider
the newly fitted FWHM(\hb) values to be more reliable than those
presented in S04, although the differences are small and the main
change is the inclusion of an estimate on the uncertainty of the
FWHM(\hb) measurement (see below). Notable exceptions are six cases
(UM\,632, 2QZ\,J231456.8-280102, [HB89]\,2254+024, SBS\,1425+606,
UM\,642 and SDSS J024933.42-083454.4) where the difference in \mbh\ is
of order $\sim$2--3 due to the change in the measured FWHM(\hb). The
total sample consists of 44 sources covering the luminosity range
$10^{45.2}<\lambda L_{\lambda}(5100\mbox{\AA})<10^{47.3}$ \ergs. This
represents the AGNs luminosity function from about a factor of 2 below
the top to a factor of $\sim 200$ below the top.

The main uncertainties on the measured masses are due to uncertainties
in FWHM(\hb). To estimate this, we divided all fits into three
categories reflecting their quality. Most of the sources show
symmetrical lines and the \ion{Fe}{2} complex is easy to model and
deconvolve. The assigned uncertainty on FWHM(\hb) in this case is
10\%. Some sources have adequate S/N ratios yet the broad line
profiles are somewhat irregular and the FWHM is more difficult to
constrain. The uncertainty in this case is estimated to be
20\%. Finally, in those cases showing asymmetric profiles, difficult
to model \ion{Fe}{2} lines and poorer S/N, we assigned an uncertainty
of 30\% on FWHM(\hb). While This procedure is somewhat subjective, we
have no better way to quantify the fitting process. We consider those
uncertainties conservative and note that they translate to relative
errors on the mass determination of \hbox{20--60\%}. None of the new
GNIRS objects is assigned the highest uncertainty of 30\% but six of
the S04 sources fall into this category.

We also used the SDSS spectra to measure $\lambda
L_{\lambda}$(1450\AA) and FWHM(\civ) for the 15 new sources. The
luminosity is available directly from the observed continuum flux and
the FWHM is obtained by fitting two Gaussians to the line
profile. Similar information is available in S04 for 27 of the 29
sources in their sample.

To summarize, our sample contains almost all sources at $z>2$ where
\mbh\ was obtained using the most robust and reliable method (the \hb\
method). Much larger uncertainties are associated with the \civ\
method (see detailed discussion below), the one used in almost all
other $z>2$ studies. Thus, we believe that our data set is the most
suitable and most accurate to address the issues of BH growth and the
distribution of \mbh\ and \Ledd\ at those redshifts.

\section{Discussion}
\label{discussion}

\subsection{Luminosity and \Ledd\ in High-Redshift AGNs}
\label{L_and_Ledd}

\begin{figure}
\plotone{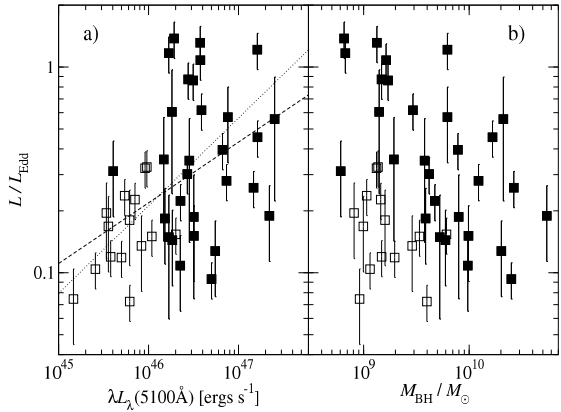}
\caption{$L/L_{\rm Edd}$ vs. $L_{5100}$ ({\it left}) and $M_{\rm BH}$
  ({\it right}) for the entire sample of 44 luminous, high-redshift
  AGNs. GNIRS (S04) sources are marked with open (full) symbols. The
  dashed and dotted lines mark two best-fit lines obtained with two,
  somewhat different statistical methods with slopes 0.29 and 0.43.}
\label{fig:Ledd_L}
\end{figure}

We have looked for correlations between \Ledd\ and various other
properties of the 44 high-redshift sources in our sample.  The
correlation with \Lop\ is shown in Fig.~\ref{fig:Ledd_L}a. The two
sub-samples are shown with different symbols emphasizing the
difference in the mean luminosity. Standard regression analysis
indicates a significant correlation with a Pearson correlation
coefficient is 0.45 ($p=2 \times 10^{-3}$) and a Spearman-rank
correlation coefficient of 0.46 ($p=2\times 10^{-3}$).  We also
checked the correlation of \Ledd\ with \mbh. The diagram is shown in
Fig.~\ref{fig:Ledd_L}b and exhibits a large scatter and no apparent
correlation. In particular, the scatter in \Ledd\ (about a factor of
10) is similar for BHs of all masses.

We have repeated the analysis using, this time, the Bentz et
al. (2007) expression for estimating \mbh.  The correlation of \Ledd\
with \Lop\ is even stronger and, again, there is no correlation of
\Ledd\ with \mbh.  As explained, all \mbh\ values calculated in this
way are smaller by factors of 1.2--1.9, depending on their
luminosity. As a result, all values of \Ledd\ are larger by similar
factors. This results in some sources with \Ledd$\sim 3$.  We suspect
that the extremely large accretion rates may not be physical but given
the method uncertainty, and the extrapolation beyond the K05
luminosity range, we cannot rule them out.  On the other hand, the
deduced \mbh\ in this case is smaller, on average by a factor 1.5,
which may be more consistent with the lack of very high mass BHs at
low redshifts (but note that even the Bentz et al. relationship gives
several cases with $M_{\rm BH} >10^{10}$\,\Msun).

The slopes of the above correlations depend on the statistical method
used and are not too different from the one expected from a case where
FWHM(\hb) is independent of \Lop\ (Eq.~\ref{eq:M_L}). Indeed, there is
no correlation between FWHM(\hb) and \Lop\ in our sample. This point
requires some explanation. Single-epoch mass determination provides
reliable BH mass estimates only because of a (yet to be explained)
scaling of the BLR size with source luminosity (discovered by
reverberation mapping) and the virial motion of the BLR gas. Given
this scaling, a complete and unbiased sample must also show some
dependence of the mean gas velocity on source luminosity, depending on
the distribution of \mbh\ in the sample. This is not observed in the
sample at hand. It may reflect the incompleteness of the sample, its
small size, or the real \mbh\ distribution. Thus, the above
correlations are not, by themselves, very important. The more
significant finding is the presence of a large number of very massive
BHs, at high redshifts, with \Ledd\ considerably smaller than
unity. In this respect, there is no difference between the two
redshift groups presented here.  As discussed below, this is relevant
to the question of BH growth in the early universe.

The recent work of Kollmeier et al. (2006; hereafter K06) includes a
systematic study of BH masses and accretion rates in a sample of 407
AGNs covering the redshift range of 0.3--4. These authors suggest a
very narrow range of \Ledd, at {\it all} luminosities and redshifts,
consistent with $\log$\,(\Ledd)=$-0.6\pm0.3$. The range is even
smaller (0.28 dex) for 131 high luminosity, high redshift ($z>1.2$ by
their definition) sources that are more relevant to the present
discussion.  The paper suggests that the intrinsic distribution in
\Ledd\ is even narrower (0.24 dex for the high-$z$ high-$L$ subgroup)
and much of the observed scatter is due to uncertainties in BH mass
determination and bolometric correction.

Our sample contains a similar number of sources to K06 in the 2.1--3.5
redshift range. Moreover, the number of K06 sources in the redshift
and \mbh\ range of our sample is much smaller (only 19 compared with
our 44). Thus our sample is more suitable, in terms of number of
sources, to address the issue of the \Ledd\ distribution in the
population of high redshift large BH mass AGNs. We find a broader
range in \Ledd\ compared with the various sub-groups presented in
K06. For example, 90\% of the 44 sources are found in the accretion
rate interval $0.08<$\Ledd$<1.5$. It is not entirely clear what is the
source of the difference between our results and those of K06. It may
be related to the large uncertainty, and perhaps even a bias, in the
method they used to determine \mbh\ (most masses in their high-$z$
high-$L$ group and all masses for $z>2$ sources were determined with
the \civ\ method; see also comment on \Ledd\ measured this way in
\S~\ref{C4}). The K06 sample is flux limited and thus more complete
than ours. However, as explained, the number of sources in individual
mass and redshift bins are extremely small. These conclusions do not
change when using the Bentz et al. (2007) \mbh\ estimate since the
discrepancy is mostly due to the range in \Ledd\ which is even larger
when this method is used.

\begin{figure}
\epsscale{1.2}
\plotone{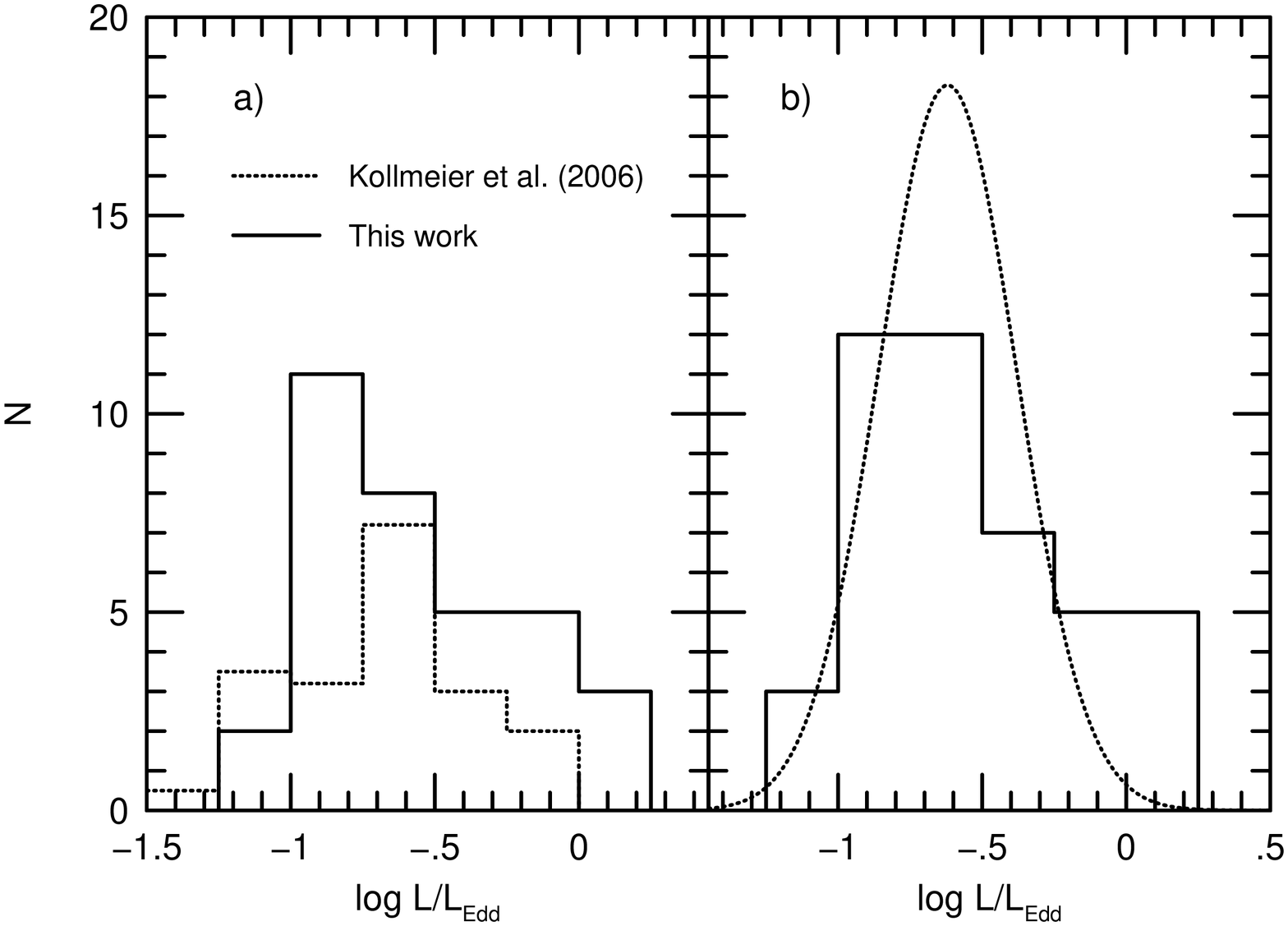}
\caption{{\it Left:} Histograms of \Ledd\ values as obtained from K06
  (Fig.~11, for $2<z<3$ sources with $M_{\rm BH}=10^{9-10}$\,\Msun,
  dotted line) and the present work (solid line, $M_{\rm
    BH}=10^{8.9-10}$\,\Msun, using $f_L=7$). Note that we directly
  copied the data from the K06 paper, thus, the somewhat different
  bolometric correction used by those authors would shift their
  histogram to the left by about 0.1 dex. {\it Right:} A comparison of
  the log(\Ledd) distribution in our sample with the favored K06 model
  (dashed line).}
\label{fig:kollmeier}
\end{figure}

To illustrate the above points, we show in Fig.~\ref{fig:kollmeier}a
one of the K06 histograms (their Fig.~11, panel with $2<z<3$ and
\mbh$=10^{9-10}$) alongside the 34 objects we observed within a
similar mass and redshift range. A visual inspection shows the
broader distribution of \Ledd\ in our sample (note that the K06
distribution should be shifted to the left by about 0.1 dex due to the
different bolometric correction used in their work).

A more quantitative test can be made by comparing our sample with the
entire high-luminosity high-redshift sub-sample of 131 objects from
K06 (Table 1), despite of the different range in $z$ and \mbh. The
overlap between the two is large but the mean luminosity and BH mass
in our sample is somewhat larger. According to K06, log(\Ledd) for
this group is well fitted with a log-normal distribution which is
centered at $-0.52$ and has a measured (model) standard deviation of
0.28 (0.24). As explained, the mean should be shifted to $-0.62$ to
allow for the somewhat larger bolometric correction used by K06.  Our
sample of 44 sources shows a similar mean $(-0.56)$ but a considerably
larger scatter of $\sim 0.35$. It also shows a \textit{positive}
skewness of $\sim 0.47$, in comparison with the K06 value of $-0.02$.
This implies that sources with larger values of \Ledd\ are more
abundant in our sample. All these points are illustrated in
Fig.~\ref{fig:kollmeier}b where we compare the histogram of our
measured values of \Ledd\ with the favored K06 distribution for the
131 high-$z$ high-$L$ sources.

We attempted to verify the null hypothesis that the values of \Ledd\
in our sample are indeed drawn from a log-normal distribution.  A
two-tailed Kolmogorov-Smirnov test over a large range of possible mean
and standard deviation gives inconclusive results. The highest
probability case has a mean log(\Ledd)$=-0.59$ and $\sigma=0.37$. The
$p$-value of this case is 0.92 (i.e. the probability of such a
log-normal distribution in our sample is 92\%).  We also find that a
K06 distribution (log(\Ledd)$=-0.62 \pm 0.24$ given our bolometric
correction) can be rejected at the 92\% level. Finally, we tested the
suggestion of a very large mean \Ledd, close to unity, in a sample
like ours. A series of Kolmogorov-Smirnov tests show that a log-normal
distribution of any width can be rejected at the 99\% confidence level
for all cases where the mean log(\Ledd) is $-0.25$ or larger. As shown
in \S~\ref{comparison}, this is relevant for the comparison with BH
evolution models.

To conclude, our sample of high luminosity high redshift sources seems
to be characterized by a similar mean but a somewhat broader
distribution in \Ledd\ compared with the similar properties
sub-samples in K06. Clearly some of the differences may be related to
the way we chose our sample and in particular the fact that it is not
a real flux limited sample. This prevents us from reaching firm
conclusions about the source of the difference at this stage.
However, the method used here is, in our opinion, the best for
measuring \mbh\ and is preferable to the K06 method (see \S~\ref{C4}).
Given all this, we suggest that the distributions of our 44 sources
shown in Fig.~\ref{fig:kollmeier} represent the AGN population, over
this redshift and \mbh\ range, in the best way. The large range in
\Ledd\ (a factor of $\sim 20$) is probably real and is likely to
represent correctly the intrinsic properties of the AGN population
analyzed here.

\subsection{Comparison with BH Evolution Models}
\label{comparison}

The cosmic evolution of BHs can be modeled by using large, recently
observed AGN samples. This requires a combination of the \hbox{X-ray}
and optical luminosity functions (LFs) since the first is more
complete yet the latter probes much deeper at high redshifts. Of the
numerous papers discussing such models we focus on the works of
Marconi et al. (2004), Merloni (2004), Hopkins et al. (2006) and
Volonteri et al. (2006).

Marconi et al. (2004) and Merloni (2004) address the growth rate of
active BHs assuming a constant accretion rate, \Ledd=1. Both papers
suggest faster growth rate of very massive BHs at earlier epochs in
order to fit the observed LFs. The more recent studies of Hopkins et
al. (2006) and Volonteri et al. (2006) include estimates of the
typical \Ledd\ required to match the redshift-dependent LFs. According
to the modeling of Hopkins et al., history of accretion and the
observed properties are not identical since the low accretion rate
phase is hardly observed due to obscuration. In particular, the large
mass very high luminosity AGNs are observable only close to their peak
luminosity and accretion rate.  This results in a narrow accretion
rate range around \Ledd=1, consistent with the Marconi et al. (2004)
assumption. Volonteri et al. (2006) used a combination of the Hopkins
et al. results with theoretical merger rates expected in cold dark
matter scenarios. The accretion rate in their models is changing with
redshift with $0.3 \le$\Ledd$\le 1$ for $3 \le z \le 6$.

The typical uncertainty on the values of \Ledd\ in our sample is a
factor of $\sim 2$ and hence the mean value is not significantly
different from the above predictions with their own large
uncertainties. However, the trend seems to be different and points in
a different direction. In particular, the theoretical requirement of
\Ledd$\simeq 1$ at high redshift and large BH mass is not seen in many
of our sources. In fact, the range of \Ledd\ covered by our very high
luminosity AGNs, from about 0.07 to about 1.7 (with the Bentz et
al. 2007 expression the range is 0.08--3.2), is not very different
from the range typically observed in low-redshift AGN samples that are
2--3 orders of magnitude less luminous. Thus, it seems that some large
mass high-redshift active BHs are observed far from their peak
luminosity phase which presents a challenge to current theoretical
models. Unfortunately, our sample does not go deep enough to search
for even lower \Ledd\ sources at high redshifts and the K06 data does
not help either since all their $z>2$ mass measurements are based on
the \civ\ method.

Finally, we mention several recent publications that attempt to
measure mass and accretion rates in even higher redshift
sources. Jiang et al. (2006) presented {\sl Spitzer} photometry of 13
AGNs at $z\sim 6$. The BH mass and accretion rate of four of the
sources were estimated~by the \civ\ method, and the \mbh\ and \Ledd\
values are given in their Table~3. All four BH masses are very large,
$\sim 5 \times 10^9$ \Msun\ and the \Ledd\ range is 0.5--1. However,
the bolometric correction used by these authors is based on
integrating over the entire spectrum, from \hbox{hard-X-rays} to the
far-infrared. This involves double-counting since almost half of this
emission is due to reprocessing of the primary continuum (accretion
disk and \hbox{X-ray}~source) by the dusty medium near the BH. Thus,
the ``true'' \Ledd\ values are smaller than those listed in their
paper by a factor of $\sim$2. Moreover, the BH mass estimates
are~rather uncertain because of the use of the \civ\ method (see
\S~\ref{C4}). Results of similar quality on a handful of additional
AGNs at $z\sim6$ have also appeared recently (e.g., Willott et
al. 2003; Jiang et al. 2007; Kurk et al. 2007).

\subsection{The Growth Rate and Growth Time of High Mass,
  High-Redshift BHs}
\label{t_grow}

The combination of the newly measured BH masses and accretion rates
can be used to estimate the growth times of massive BHs at $z\sim 2.3$
and $z \sim 3.4$. We follow the procedure outlined in NT07 (their
Eq.~6) to calculate $t_{\rm grow}$ assuming
\begin{equation}
  t_{\rm grow} = t_{\rm Edd} \frac {\eta /(1- \eta) }{f_L L_{5100}/L_{\rm Edd}} \log
  \left ( \frac{M_{\rm BH}}{M_{\rm seed}}\right ) \frac{1}{f_{\rm active}} \,\,  ,
\label{eq:t_grow}
\end{equation}
where $t_{\rm Edd}=3.8\times 10^8$\,yr for cosmic abundance, $\eta$ is
the accretion efficiency, and $f_{\rm active}$ is the duty cycle (the
fractional activity time) of the BH. The growth time, $t_{\rm grow}$,
is most sensitive to $\eta$ and $f_{\rm active}$ since both can vary
by large factors. The seed BH mass, $M_{\rm seed}$, may also change
over a large range but $t_{\rm grow}$ is not very sensitive to this
change. A small seed BH ($10^2-10^3$\,\Msun) can be the result of
population~III stars, at $z\sim 20$, while larger seed BHs
($10^4-10^6$\,\Msun) can be due to direct collapse at lower redshifts
(Begelman, Volonteri, \& Rees, 2006). Finally, $t_{\rm grow}$ is less
sensitive to $f_L$ because of the limited expected range of this
factor in the sample at hand.

As discussed in several publications, most recently by King \& Pringle
(2006), the value of $\eta$ depends on the direction and magnitude of
the BH spin and the angular momentum of the accretion disk. The
highest value is $\sim 0.4$ and the lowest values ($\sim 0.04$) is
obtained for retrograde accretion with a BH spin parameter of
$a=-1$. More typical values, that reflect several spin-up and
spin-down episodes with small-mass disks (see King \& Pringle 2006),
give $\eta$ in the range \hbox{0.05--0.08}. The values of $t_{\rm
  grow}$ calculated here (Table~\ref{table_properties}) are for the
case of $\eta=0.1$, $M_{\rm seed}=10^4$\,\Msun, $f_{\rm active}=1$,
and $f_L=7$. We list them as $t_{\rm grow}/t_{\rm universe}$ where
$t_{\rm universe}$ is calculated at the given redshift of the source
(using our adopted cosmology from \S~\ref{observations}).

The computed values of $t_{\rm grow}$ exceed the age of the universe
in most of the lower \Ledd\ sources. While somewhat smaller values of
$t_{\rm grow}/t_{\rm universe}$ are indeed likely, given the expected
range in $\eta$, we consider the numbers given in
Table~\ref{table_properties} as conservative lower limits since
$f_{\rm active}$ is likely to be much smaller than unity. For example,
general galaxy and BH growth considerations predict $f_{\rm active}
\ltsim 0.1$.  This can be estimated from number counts of AGNs and
galaxies at large redshifts, from estimates of typical time scales for
powerful starbursts, and from models of BH and galaxy
evolution. Another estimate is obtained by assuming that the ratio of
BH mass to the host galaxy mass is similar to the typical value
observed in the local universe (about 1/700). This translates, for
most objects in our sample, to a host mass of $10^{12}$\,\Msun\ or
larger. To build up such a large stellar mass would require continuous
star formation, from very early times, at a rate of $\sim
300$\,\Msun/yr, or a much higher star formation rate with a
star-formation duty cycle which is less than unity. Given the much
faster growth of BHs by accretion, their duty cycles would be
considerably shorter. While some very different scenarios cannot be
excluded on the basis of current observations, they are definitely not
in accord with present day galaxy-formation models.  Anti-hierarchal
BH growth models would give faster BH growth at high redshifts with a
larger BH-to-galaxy mass ratio. It remains to be seen whether any such
model is consistent with $f_{\rm active} \sim 1$ all the way to
redshift 3 for BHs like the ones in our sample.

Given our more plausible assumption of $f_{\rm active} \ll 1$ for
$z=2-3$ very large mass BHs, all 15 new sources presented here, and
several of the remaining more luminous objects, require too much time
to grow to their observed size. This problem was not noted in earlier
works since most of them assumed \Ledd$\sim 1$ and hence $t_{\rm
  grow}$ an order of magnitude shorter. The problem is most severe for
the lowest accretion rate sources. The higher accretion rate sources
(most of the 29 objects from S04) have just enough time to explain
their mass if $f_{\rm active} \sim 1$. Thus, our new measurements, and
general considerations, indicate that some very massive BHs had just
enough time to grow to their measured size at redshifts $\sim2.3$ and
$\sim3.4$. However, for all sources with \Ledd$ \ltsim 0.3$, there was
not enough time at redshift $\sim3.4$, and even at redshift $\sim2.3$,
to grow to their observed size. This problem is more severe if we
adopt the Hopkins et al. (2006) set of models where much of the BH
growth is during times where \Ledd\ is smaller than the time where the
object is not obscured.

We conclude that a significant fraction of the sources in our sample
must have had at least one previous episode of faster growth, probably
with \Ledd$\sim 1$ and at $z\gtsim2-3$, in order to explain their BH
mass. Given this, many extremely large mass BHs at redshifts 2 and 3
are in the process of their second and perhaps even third or fourth
episode of activity. This does not necessarily apply for small-mass
BHs at those redshifts that may have a different growth pattern.

\subsection{\civ\ as a BH-Mass Indicator}
\label{C4}

As noted above, the \civ\ method for estimating BH masses is
problematic since \civ\ lines are known to show unusual profiles in
many cases, including an enhanced blue wing and a large wavelength
shift relative to the systemic velocity. This issue has been
discussed, extensively, in numerous papers. The work of Baskin \& Laor
(2005) clearly shows the differences between the \hb\ and \civ\ line
profiles and the problematics of FWHM(\civ) as a virial-motion
indicator (see also the recent Shang et al. 2007 work). However,
according to Vestergaard \& Peterson (2006), a proper calibration of
FWHM(\civ) and the UV continuum provides an adequate replacement for
the \hb\ method. K06 adopted this view and used the \civ\ method to
estimate BH masses in many high-redshift sources.

\begin{figure}
\plotone{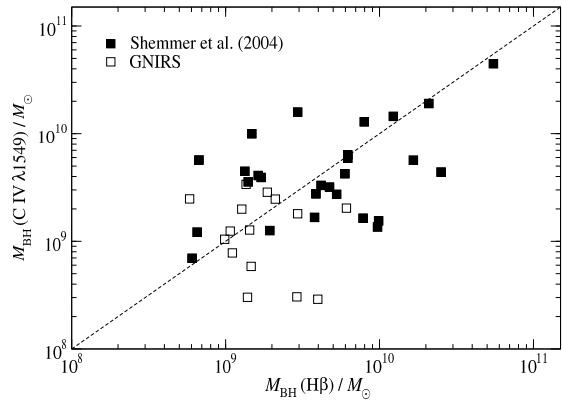}
\caption{BH mass calculated with the \hb\ method vs. the one obtained
  from the combination of FWHM(\civ) and the UV continuum. Symbols are
  as in Fig.~\ref{fig:Ledd_L}. The dashed line marks a 1:1
  correspondence, to guide the eye.}
\label{fig:M_hb_M_civ}
\end{figure}

Our combined sample includes 44 extremely luminous AGNs with FWHM(\hb)
and \Lop\ values for all. We have compiled $\lambda
L_{\lambda}$(1450\AA) and FWHM(\civ) for 42 of these and are thus in a
position to compare the \hb\ and \civ\ methods in the high luminosity
range. We first checked \Lop\ vs. $\lambda L_{\lambda}$(1450\AA) for
all sources where we have independent luminosities for both (i.e.,
excluding the six sources from S04 for which optical luminosities were
estimated from their UV luminosities). We find a very significant, low
scatter correlation which is consistent with $L_{\nu} \propto
\nu^{-0.5}$. The diagram (not shown here) suggests that the two
estimates of the source luminosity can be used, interchangeably, with
a different normalization, for estimating the BLR size. We then
obtained BH masses by using the two methods; the \hb\ method
(Eq.~\ref{eq:M_L}) and the \civ\ method with the expression given in
Vestergaard \& Peterson (2006). The two mass estimates are shown in
Fig.~\ref{fig:M_hb_M_civ}. The figure is a complete scatter diagram
with no apparent correlation or trend. While the median (0.89) and the
mean (1.49) \mbh(\civ)/\mbh(\hb) ratio are not far from unity, the
scatter is about $\pm 0.3$ dex.

Repeating the same analysis by using the Bentz et al. (2007)
expression, gives also a scatter diagram with no significant
correlation. In this one, the median and the mean for
\mbh(\civ)/\mbh(\hb) are 1.3 and 2.2, respectively and the scatter is
about 0.4 dex. This may provide a slight indication that the slope of
$\alpha=0.65$ used in our work is preferable to the one used by Bentz
et al. (2007) since, using this slope, the agreement between the
median \mbh(\civ) and the median \mbh(\hb) is improved.

The lack of correlation between \mbh(\hb) and \mbh(\civ) suggests
that, while BH mass estimates based on the \civ\ line properties
roughly follow the increase in source luminosity, their use is rather
limited and highly uncertain, in the high luminosity and/or high
redshift range. This conclusion is independent of the slope used to
obtain \mbh(\hb). As argued earlier, some of the differences found
between our \Ledd\ distribution and the one presented in K06 are
likely to be the result of their use of the \civ\ method. In fact, we
notice a complete lack of correlation between the \Ledd\ values
derived with the \hb\ method and those derived with the \civ\ method.
Given all this we suspect that the unexpected trend found by K06 when
comparing the various line-based mass measurements (see their Fig.~4
and related discussion) is due to the problematic \civ\ method.

\section{Conclusions}
\label{conclusions}

We have presented new, high quality observations of the \Hb\ region in
15 high luminosity, high-redshift AGNs. We used the new data, in
combination with our earlier observations of 29 sources, to compile a
sample of 44 high luminosity AGNs where the BH mass estimates are all
based on the \hb\ method. The main findings are:
\begin{enumerate}
\item Our sample is the largest of its type, yet, and can therefore be
  used to investigate, in a more reliable way, the important
  correlations between source luminosity, BH mass and accretion
  rate. Such measurements at high redshifts can be combined with
  similar measurements at $z<0.75$ to follow BH growth and AGN
  evolution through time.
\item There is a significant correlation between \Lop\ and \Ledd\ but
  no correlation between \mbh\ and \Ledd.  These results do not depend
  on the exact slope used to derive \mbh.
\item Assuming the distribution in \Ledd\ found here represents the
  high-redshift high luminosity AGN population, we find a significant
  fraction of sources with small ($\le 0.2$) \Ledd. Current
  theoretical models predict only very few such sources among
  high-redshift AGNs.
\item Low accretion rate BHs represent a real challenge to BH growth
  scenarios. A possible way out is to assume that in all such cases,
  there was at least one earlier ($z>2.3$ or $z>3.4$) episode of BH
  growth with a higher accretion rate.
\item We present new evidence for the large uncertainty and probably
  systematic error associated with the use of the \civ\ method for
  estimating \mbh\ and hence \Ledd\ in high-redshift, high luminosity
  AGNs.
\end{enumerate}

\acknowledgements We thank J.~M. Wang for useful discussions, Dovi
Poznansky for assistance with the synthetic photometry and an
anonymous referee for useful suggestions. Funding for this work has
been provided by the Israel Science Foundation grant 232/03 (HN and
BT), by the Jack Adler chair of Extragalactic Astronomy at Tel Aviv
University (HN) and by Project Fondecyt \#1040719 (PL). This work is
based on observations obtained at the Gemini Observatory, which is
operated by the Association of Universities for Research in Astronomy,
Inc., under a cooperative agreement with the NSF on behalf of the
Gemini partnership: the National Science Foundation (United States),
the Particle Physics and Astronomy Research Council (United Kingdom),
the National Research Council (Canada), CONICYT (Chile), the
Australian Research Council (Australia), CNPq (Brazil) and CONICET
(Argentina).

\end{document}